\documentclass[12pt,preprint]{aastex}
\shorttitle{Comparing P-stars with Observations}
\shortauthors{P.Cea}

\begin{document}

   \title{COMPARING P-STARS WITH OBSERVATIONS}
   \author{Paolo Cea}

   \affil{Dipartimento Interateneo di Fisica, Universit\`a di Bari,
              Via G. Amendola 173, I-70126 Bari\\
         \and
             INFN - Sezione di Bari, Via G. Amendola 173, I-70126 Bari \\
                \email{Paolo.Cea@ba.infn.it}
             }

\begin{abstract}
 P-stars are compact stars made of up and down quarks in $\beta$-equilibrium with electrons in a chromomagnetic
condensate. P-stars are able to account for  compact stars as well as stars with radius comparable with canonical neutron stars.
We compare p-stars with different available observations. Our results  indicate that p-stars are  able to reproduce in a natural manner several observations from isolated and binary pulsars.
\end{abstract}
\maketitle
\section{INTRODUCTION}
\label{Introduction}
In few years since their discovery~\cite{Hewish:1968} pulsars have been identified with rotating neutron stars, first predicted
theoretically by Baade \& Zwicky (1934a,b,c), endowed with a strong magnetic field~\cite{Pacini:1968, Gold:1968}.  \\
\indent
It is widely believed that there are no alternative models able to provide as satisfactory an explanation for the wide variety of pulsar phenomena as those built around the rotating neutron star model.  Nevertheless, recently we have proposed (Cea 2004a,b) a new class of compact stars, p-stars, which is challenging  the standard paradigm.  \\
\indent
P-stars are compact stars made of up and down quarks in $\beta$-equilibrium with electrons in a chromomagnetic condensate. In our previous studies (Cea 2004a,b), we find that p-stars are able to account for  compact stars,  stars with radius comparable with canonical neutron stars, as well as super massive compact objects. Moreover, we show that p-stars once formed are absolutely stable. The cooling curves of p-stars compare rather well with observations. We also suggested that p-matter produced at the cosmological deconfinement  phase transition could be a viable candidate for baryonic cold dark matter. \\
\indent
In addition, in Cea (2006) we  discuss p-stars endowed with super strong dipolar magnetic field. We find that soft gamma-ray repeaters and anomalous X-ray pulsars can be understood within our theory. In particular, we point out that within our p-star model there is a quite natural mechanism to account for the generation of dipolar surface magnetic fields up to $10^{16} \, Gauss$.  We succeed in obtaining a well defined criterion to distinguish rotation powered pulsars from magnetic powered pulsars. We show that glitches, that in our theory  are triggered by magnetic dissipative effects in the inner core, explain both the quiescent emission and burst activity in soft gamma-ray repeaters and anomalous X-ray pulsars. We are able to account for braking and normal glitches observed in  soft gamma-ray repeaters and anomalous X-ray pulsars.  We discuss and explain the observed anti correlation between hardness ratio and burst intensity. Within our p-star theory we are able to account quantitatively for  light curves from both gamma-ray repeaters and anomalous X-ray pulsars. \\
\indent
It is worthwhile to briefly discuss the theoretical foundation of our proposal. Indeed,  quite recently, the QCD vacuum was probed by means of an external constant abelian chromomagnetic field on the lattice~\cite{cea:2003,cea:2005}. We found that by increasing the strength of the applied external field the
deconfinement temperature decreases towards zero. In other words, there is a critical field $gH_c$ such that for $gH>gH_c$ ($g$ is the color gauge coupling and $H$ is the strength of the  chromomagnetic field directed along the third direction in color space) the gauge system is in the deconfined phase (the color Meissner effect).  As a consequence, we see  that there is an intimate connection between abelian chromomagnetic fields and color confinement. The existence of a critical chromomagnetic field  is compatible with the QCD vacumm behaving like a disordered chromomagnetic condensate, for strong enough chromomagnetic field strengths enforce long range colour order thereby
destroying confinement.  In a seminal paper, R. P. Feynman (1981)  argued that QCD in two spatial dimensions   the confining vacuum at large
distances is a chromomagnetic condensate disordered by the gauge invariance. The confinement of colors comes from the
existence of a mass gap and the absence of colour long range order. As a matter of fact, we were able to show that, indeed, the
QCD vacuum in three spatial dimensions does display at large distance a mass gap and no colour long range order and
that the vacuum displays the color Meissner effect. We find that chiral symmetry breaking for quarks in the fundamental
color representation is an inevitable consequence of the disordered chromomagnetic condensate. Finally, we argue that the
deconfined vacuum behaves more like a correlated liquid than an ideal gas, in accordance with  recent results from heavy ion experiments.
This, in turns, leads us to the conclusion that the deconfined QCD vacuum is characterised by long-range chromomagnetic correlations  and whence  p-matter, namely almost massless up and down quarks immersed in a chromomagnetic condensate, must be quite close to the true QCD deconfined state. In absence of strong gravity effects, the chromomagnetic condensate in p-matter is of order of the critical field strength, which turns out to be $\sqrt{gH_c} \,   \simeq  \, 1.0 \; Gev$~\cite{cea:2003,cea:2005}. \\
\indent
Note that we are using natural units where  $\hbar \; = \; c \; =  \; 1$. In this units $g$ is dimensionless and  $\sqrt{gH}$ has dimension of energy. To switch to cgs units it suffices to replace $gH $ with  $\hbar  c \; gH$. \\
\indent
On the other hand,  when including the effects of gravity it turns out that p-matter gives rise to compact stars (p-stars), which  are more stable than neutron stars~\cite{cea:2004a} whatever the value of  $\sqrt{gH}$.  Therefore, we have in general that in p-stars the chromomagnetic condensate must satisfy the constraint:
\begin{equation}
\label{1.1}
\sqrt{gH} \; \leq  \sqrt{gH_c} \;   \simeq  \; 1.0  \; Gev \; \; .
\end{equation}
We see, then, that the chromomagnetic condensate in p-stars acts like a dynamical effective bag constant which can be varied according to Eq.~(\ref{1.1}). The actual value of the chromomagnetic condensate depends only on the central density of the compact star. In this way,  p-stars are able to overcome the gravitational collapse irrespective to the stellar mass.  Indeed, as discussed in Cea (2004a),   our p-stars do not admit the existence of an upper limit to the mass of a completely degenerate configuration. In other words, our peculiar equation of state of degenerate up and down quarks in a chromomagneticn condensate allows the existence of finite equilibrium states for stars of arbitrary mass. To see this, we note that
on dimensional ground from the Tolman-Oppenheimer-Volkov equation~\cite{shapiro:1983} we get:
\begin{equation}
\label{1.2}
 M \; = \; \frac{1}{G^{3/2} gH} \; \;
 f(\overline{\varepsilon}_c) \;
\; \; \; \; \; R \; = \; \frac{1}{G^{1/2} gH} \; \;
g(\overline{\varepsilon}_c) \; ,
\end{equation}
where $\overline{\varepsilon}_c \, = \, \varepsilon_c/(gH)^2 $ and ${\varepsilon}_c $  is the stellar central density. As
a consequence we obtain:
\begin{equation}
\label{1.3}
\frac{2 \; G \; M}{R} \; = \; 2 \;
\frac{f(\overline{\varepsilon}_c)}{g(\overline{\varepsilon}_c)} \;
\; \equiv \; h(\overline{\varepsilon}_c).
\end{equation}
\indent
For reader's convenience, it is worthwhile to rewrite the previous equations in cgs units. We get:
\begin{equation}
\label{1.4}
 M \; = \; \frac{\hbar^{1/2} \; c^{9/2}}{G^{3/2} gH} \; \;
 f(\overline{\rho}_c) \;
\; \; \; \; \; R \; = \; \frac{\hbar^{1/2} \; c^{5/2}}{G^{1/2} gH} \; \;
g(\overline{\rho}_c) \; ,
\end{equation}
where $\overline{\rho}_c \, = \, (\hbar \; c^{3} \; \rho_c)/(gH)^2 $ and ${\rho}_c $  is the stellar central density in $gr/cm^3$, 
\begin{equation}
\label{1.5}
\frac{2 \; G \; M}{R \; c^2} \; = \; 2 \;
\frac{f(\overline{\rho}_c)}{g(\overline{\rho}_c)} \;
\; \equiv \; h(\overline{\rho}_c).
\end{equation}
\indent
From Eqs.~(\ref{1.2}) and (\ref{1.3}) we see that by decreasing the strength of the chromomagnetic condensate we increase the mass and radius of the star, while the ratio $\frac{2 \; G \; M}{R}$ depends only on  $\overline{\varepsilon}_c $. It
turns out~\cite{cea:2004a} that the function $h(x)$ defined in Eq.~(\ref{1.3}) is less than 1 for any allowed values of $\overline{\varepsilon}_c$.  By varying the chromomagnetic condensate strength in the allowed range, the M-R p-star curves span the allowed region in the M-R plane. Indeed, a clear observational evidence in favour of p-stars would be the detection of pulsars with mass above $3 \; M_{\bigodot}$.
\\
\indent
In the present paper we focus on few selected topics with the aim to furnish further compelling evidences in support of our proposal. In section~\ref{masses} we compare recent determination of masses and radii of isolated and binary pulsars with our p-star model. In section~\ref{cooling}, after a brief review of cooling in p-stars, we compare our theoretical cooling curves with several observational data. Finally, we draw our conclusions in section~\ref{conclusion}.
\section{MASSES and RADII}
\label{masses}
In a recent analysis of the low-mass X-ray  binary pulsar  EXO 0748-676 observational  data  \"Ozel (2006) concluded that the determination of the mass and radius of this pulsar appears to rule out all the soft equations of sate of neutron-star matter. Indeed, multiple phenomena have been observed from this low-mass X-ray binary that can be used to determine uniquely the mass and radius. \\
\indent
The three quantities used in \"Ozel (2006) are the Eddington limit $F_{Edd}$, the gravitational redshift $z$, and the ratio $F_{cool}/\sigma T_c^4$.  Following \"Ozel (2006), we have:
\begin{equation}
\label{2.1}
 F_{Edd} \; = \; \frac{1}{4 \pi D} \; \frac{4 \pi G M}{\kappa_e} \; \left [ 1 \, - \, \frac{2 G M}{R} \right ]^{\frac{1}{2}}
 \; \; ,
\end{equation}
\begin{equation}
\label{2.2}
z \; = \;  \left [ 1 \, - \, \frac{2 G M}{R} \right ]^{- \frac{1}{2}}
 \; \; ,
\end{equation}
\begin{equation}
\label{2.3}
 \frac{F_{cool}}{\sigma T_c^4}  \; = \; f_{\infty}^{-4}  \; \frac{R^2}{ D^2} \;  \left [ 1 \, - \, \frac{2 G M}{R} \right ]^{-1}
 \; \; .
\end{equation}
Here and in the following we shall adopt natural units where $\hbar \; = \; c \; = \; k_B \; = \; 1$. \\
\indent
In Eqs.~(\ref{2.1}), (\ref{2.2}), and  (\ref{2.3}) $G$ is the gravitational constant, $M$ and $R$ are the stellar mass and radius, $D$ the distance to the source, and 
$\kappa_e$ is the electron scattering opacity:
\begin{equation}
\label{2.4}
\kappa_e  \; \simeq  \; 0.2 \; ( 1 \; + \; X) \; cm^2 \, g^{-1}   \;  \;  ,
\end{equation}
$X$ being the hydrogen mass fraction of the accreted material. Finally, $ f_{\infty}$ is the colour correction factor which relates the colour temperature $T_c$ to the effective temperature $T_{eff}$ of the star. \\
\indent
\"Ozel pointed out that the following  parameterization of the colour correction factor:
\begin{equation}
\label{2.5}
f_{\infty} \; \simeq  \; 1.34 \; + \; \left ( \frac{1 +X}{1.7} \right )^{2.2} \;   \left [ \frac{(T_{eff}/10^7 \, {}^\circ K)^4}
{g/(10^{13} \, cm \, s^{-2})}  \right ]^{2.2} \;  \;  ,
\end{equation}
leads to an accurate description  of the results from model atmosphere calculations   \cite{Madej:2004}.  \\
\indent
Remarkably,  Eqs.~(\ref{2.1}),  (\ref{2.2}), and  ( \ref{2.3}) can be solved to uniquely determine the stellar mass, radius, and 
distance~\cite{Ozel:2006}:
\begin{equation}
\label{2.6}
 M  \; = \; \frac{ f_{\infty}^4 }{4 \kappa_e G} \;  \;  \frac{[ 1 \, - \, (1 + z)^{-2} ]^2}{(1 + z)^3} \; \;   \frac{F_{cool}}{\sigma T_c^4} \; \;  \frac{1}{F_{Edd}} \; \; ,
\end{equation}
\begin{equation}
\label{2.7}
R \; = \;  \frac{ f_{\infty}^4 }{2 \kappa_e} \;  \;  \frac{ 1 \, - \, (1 + z)^{-2}}{(1 + z)^3} \; \;   \frac{F_{cool}}{\sigma T_c^4} \; \;  \frac{1}{F_{Edd}}  \;  \;  , 
\end{equation}
\begin{equation}
\label{2.8}
D   \; = \; \frac{ f_{\infty}^2 }{2 \kappa_e} \;  \;  \frac{ 1 \, - \, (1 + z)^{-2}}{(1 + z)^3} \; \;  \left [ \frac{F_{cool}}{\sigma T_c^4} \right ]^{\frac{1}{2}}  \; \;  \frac{1}{F_{Edd}}  \;  \;  .
\end{equation}
The value of the redshift has been reported in Cottam et al. (2002):
\begin{equation}
\label{2.9}
z \; = \; 0.35 \;  \;  .
\end{equation}
Moreover, the Eddington limit luminosity  $F_{Edd}$ has been obtained in \"Ozel (2006) by averaging the values determined with RXTE~\cite{Wolff:2005} and EXO SAT~\cite{Gottwald:1986}:
\begin{equation}
\label{2.10}
F_{Edd} \; = \; 2.25 \; \pm 0.23 \; 10^{-8} \; erg \, cm^{-2} \, s^{-1}  \;  \;  .
\end{equation}
Finally, the ratio $F_{cool}/\sigma T_c^4$ has been inferred from the RXTE data~\cite{Wolff:2005}:
\begin{equation}
\label{2.11}
 \frac{F_{cool}}{\sigma T_c^4} \; = \; 1.14 \; \pm 0.10  \;  \left ( km/kpc \right )^2  \;  \;  .
\end{equation}
Using the values in Eqs.~(\ref{2.9}),  (\ref{2.10}), and  (\ref{2.11}) we easily obtain from   Eqs.~(\ref{2.6}),  (\ref{2.7}), and  (\ref{2.8}):
\begin{equation}
\label{2.12}
 \frac{M}{M_{\bigodot}}  \; = \; \left [ 0.2009 \; \pm \; 0.0271 \right ] \; \; \frac{ f_{\infty}^4 }{\kappa_e } \;  \;  ,
\end{equation}
\begin{equation}
\label{2.13}
R \; = \;  \left [ 1.315 \; \pm \; 0.177 \right ] \;  km \;  \frac{ f_{\infty}^4 }{\kappa_e } \;  \;  ,
\end{equation}
\begin{equation}
\label{2.14}
D   \; = \; \left [ 1.232 \; \pm \; 0.137 \right ] \;  kpc \;  \frac{ f_{\infty}^2 }{\kappa_e } \;  \;  .
\end{equation}
The main uncertainty that affects the mass, radius, and distance of the binary pulsar resides in the hydrogen mass fraction $X$ of the accreting material and the colour correction factor $f_{\infty}$. \\
\indent
Following \"Ozel (2006), if  we assume for the colour correction factor:
\begin{equation}
\label{2.15}
f_{\infty} \; \simeq  \; 1.37 \;  \;  ,
\end{equation}
and using the extreme value of hydrogen mass fraction $X \, \simeq \, 0.70$, we get :
\begin{equation}
\label{2.16}
 \frac{M}{M_{\bigodot}}  \; = \; 2.10  \; \pm \; 0.28  \;  \;  ,
\end{equation}
\begin{equation}
\label{2.17}
R \; = \; 13.62 \; \pm \; 1.84  \;  km \;  \;  \;  .
\end{equation}
\begin{figure}[ht]
   \centering
   \epsscale{0.55}
  \plotone{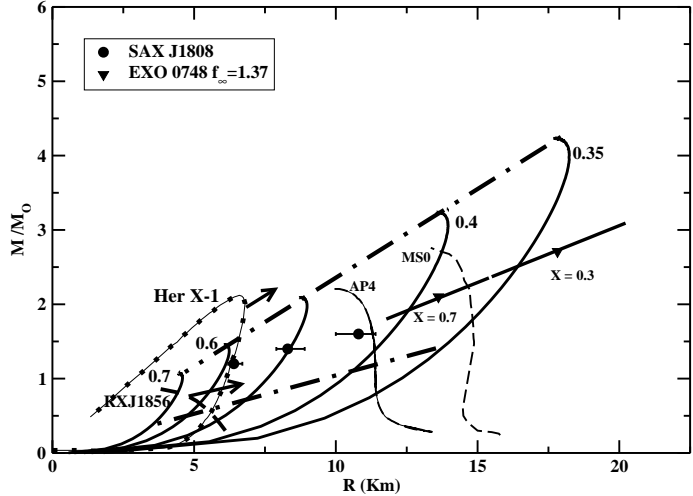}
\caption{\label{fig1}
Full lines are the gravitational mass M versus the stellar radius R for p-stars for different values of $\sqrt{gH}$ (in Gev). The  dashed lines  are the mass-radius curves for neutron stars by assuming the stiffest equation of state ($MS0$) and  a softer equation of state ($AP4$, see Lattimer \& Prakash 2001).  Full triangles correspond to the EXO 0748-676 mass-radius constraints assuming $f_{\infty}  \simeq   1.37$ for two different values of hydrogen mass fraction $X$. Full circles  are the best-fitting mass-radius values of the model 2 in Table~1 of  Poutanen \& Gierlinski (2003) for the  accreting X-ray millisecond pulsar J1808. The thick dashed line corresponds to the mass-radius curve for RXJ 1856 obtained by solving Eq.~(\ref{2.25}) with $R^\infty   \simeq 6.6 \; Km$.  The dotted line is the semi-empirical M-R curve for Her X-1 discussed in Reynolds et al. (1997) with assumed luminosity of $L_X = 3.5 \; \times \; 10^{37} \; erg \, s^{-1}$. The arrows indicate that the M-R curves for  Her X-1 and  RXJ 1856 are to be intended as lower limits. The region comprises between the two dot-dahed lines and the M-R curves with  $\sqrt{gH} = 0.6 Gev, 0.35 Gev$ is the allowed region for p-star.}
\end{figure}
On the other hand, for binary systems like EXO 0748-676 a helium-rich companion should be expected (see, for instance Lewin  1993). In this case the value  $X \, \simeq \, 0.30$ seems to be more appropriate. If this is the case, we get:
\begin{equation}
\label{2.18}
 \frac{M}{M_{\bigodot}}  \; = \; 2.72  \; \pm \; 0.37  \;  \;  ,
\end{equation}
\begin{equation}
\label{2.19}
R \; = \; 17.82 \; \pm \; 2.40  \;  km \;  \;  \;  .
\end{equation}
In Fig.~\ref{fig1} we compare Eqs.~(\ref{2.16}),  (\ref{2.17}), and Eqs.~ (\ref{2.18}),  (\ref{2.19})  with the mass-radius relation obtained within our p-star theory \cite{cea:2004a}. We can see that for both hydrogen abundances our model is able to account for the inferred values, Eqs.~(\ref{2.16}),  (\ref{2.17}) and Eqs.~(\ref{2.18}),  (\ref{2.19}), together with the constraint Eq.~(\ref{2.9}). On the contrary, the mass-radius curve for neutron stars is consistent with the determination Eq.~(\ref{2.16}) and Eq.~(\ref{2.17}) only by assuming the stiffest equation of state (labelled $MS0$ in Fig.~\ref{fig1}, see  Lattimer 2001), which corresponds to high-density neutron matter without non-linear vector and isovector interactions~\cite{Muller:1996}. Note, however, that even the stiffest high-density  neutron matter equation of state is ruled out if we consider the more realistic hydrogen mass fraction $X \simeq 0.30$. \\
\indent
Interestingly enough, we find that regardless of  the actual value of  hydrogen mass fraction the distance of the binary system is quite tightly  constrained:
\begin{equation}
\label{2.23}
 \left ( 8.87 \; \pm \; 0.99 \right ) \;  kpc \; \lesssim  \;D   \; \lesssim \; \left ( 12.38 \; \pm \; 1.39 \right ) \;  kpc \;  \;  .
\end{equation}
\indent
We see that  Fig.~\ref{fig1}  shows that EXO 0748-670 can be accounted for by our p-star model irrespective on the assumed value of the hydrogen mass fraction. \\
\indent
We see that  the stiffest equation of state for neutron-star matter is marginally  compatible with observations from EXO 
 0748-670. However,  we argue in a moment that there  is some tension  with several pulsar data which would require a
 softer equation of state.  \\
\indent
It has been proposed long time ago that the compact accreting object in the famous X-ray binary  Herculses X-1 is a
strange star \cite{Li:1995}. This proposal was based on the comparison of a phenomenological  mass-radius relation for
Herculses X-1 (see for instance Shapiro \& Teukolsky 1983) with theoretical M-R curves for neutron and strange stars. The
analysis in Li et al. (1995) has, however, been criticized by  Reynolds et al. (1997). These authors, using a new mass
estimate together with a revised distance, which leads to a somewhat higher X-ray luminosity, argued that the hypothesis
that  Herculses X-1 is a neutron star is not disproved. As a matter of fact, Reynolds et al. (1997) found that there is
marginal consistency with observations if one adopts for neutron stars a very soft equation of state. At the same time,
these authors pointed out that the hypothesis of a strange star can be ruled out since the theoretical curves no longer
intercept the observational relations within the permitted mass range.  However, it is evident from Fig.~1 in Reynolds et al. (1997) that the mass-radius constraints for Her X-1 seem to be  incompatible with the stiffest equation of  state  displayed in our Fig.~\ref{fig1} where we display  the semi-empirical M-R curve for Her X-1 discussed in Reynolds et al. (1997) corresponding to the luminosity $L_X = 3.5 \; \times \; 10^{37} \; erg \, s^{-1}$. On the other hand, the phenomenological  mass-radius relation for Herculses X-1 could be compatible with a softer equation of state like the one labelled AP4 in Fig.~\ref{fig1} (for the nomenclature  on the equations of state see  Lattimer \& Prakash 2001). \\
\indent
The millisecond pulsar SAX J1808-3658 (henceforth J1808) is one of the most studied  accreting pulsar (for instance, see Van der Klis  1995). Recently,  Poutanen \& Gierlinski (2003) studied the pulse profile of J1808 at different energies. In particular,  they derived simple analytical formulae for the light curves. By fitting the observed pulse profiles in the energy band $3-4 \; keV$ and $12-18 \; keV$ Poutanen \& Gierlinski (2003) constrain the compact star mass and radius. The best-fitting parameters for two different models are presented in Table~1 of  Poutanen \& Gierlinski (2003). It turns out that the results from the two adopted  models are quite consistent \cite{Poutanen:2003}.  In Fig.~\ref{fig1} we report the results form model 2 in Poutanen \& Gierlinski (2003). Again, we see that  the mass-radius constraints for J1808 are more consistent with a soft  equation of state for  high density neutron matter. On the other hand,  our p-star model  can easily  account for the observed mass-radius values for the  accreting X-ray millisecond pulsar J1808.  \\ 
\indent
Remarkably, our previous conclusion is confirmed by the recent analysis of the light curves of J1808 during its 1998 and 2002 outbursts~\cite{Leahy:2007}. Indeed, Leahy et al (2007) obtain that at the $3 \; \sigma$ level the radius must satisfy $ R \, < \, 11.9 \; km$  and the mass $ M \, < \, 1.56 \; M_{\bigodot}$ . \\
\indent
RXJ 1856.5-3754 (in the following RXJ 1856) is the nearest and brightest of a class of isolated radio-quiet compact stars. RXJ 1856  has been observed with Chandra and XMM-Newton~\cite{burwitz:2003} , showing that the X-ray spectrum is accurately fitted by a blackbody law. Assuming that the X-ray thermal emission  is due to the surface of the star, Burwitz et al. (2003)  found for the effective radius and surface temperature:
\begin{equation}
\label{2.24}
 R^\infty \, \simeq \, 4.4 \, \frac{d}{120 \, pc} \;  Km \; \; , \; \; T^\infty \; \simeq
 \; 63 \; eV  \; \; ,
\end{equation}
where
\begin{equation}
\label{2.25}
 R^\infty \; = \;  \frac{R}{\sqrt{ 1 \, - \frac{2GM}{ R}}} \; \; \; , \; \;  \;
 T^\infty \; = \;  T \;  \sqrt{ 1 \, - \frac{2GM}{ R}} \; \; .
\end{equation}
It should be stressed, however,  that in the observed spectrum there is also an optical emission  in excess over the extrapolated
X-ray blackbody. By interpreting the optical emission as a Rayleigh-Jeans tail of a thermal blackbody emission, one finds
that the optical and X data can be also fitted by the two  blackbody model. In this case, the spectral fit  yield an effective radius~\cite{burwitz:2003}:
\begin{equation}
\label{2.26}
R^\infty \;  \gtrsim  \; 16 \; Km \;  \frac{d}{120 \, pc} \; \; .
\end{equation}
The two blackbody model, however, does not furnish an acceptable description of the observed spectrum. Indeed,
interestingly enough, quite recently the distance measurement of RXJ 1856 has been reassessed and it is now estimated to be
at about $180 \; pc$~\cite{kaplan:2003}. Moreover, from recent parallax measurements~\cite{kaplan:2003,vanKerkwijk:2006} we infer that there is a lower limit to the distance of  RXJ 1856:
\begin{equation}
\label{2.27}
d \;  \gtrsim  \; 160  \;   pc  \; \; .
\end{equation}
In fact, Eq.~(\ref{2.27}) excludes  the two-blackbody interpretation, for  in this case from Eq.~(\ref{2.26})  we should obtain:
\begin{equation}
\label{2.28}
R^\infty \;  >  \; 21 \; Km  \;  \; ,
\end{equation}
which is too large for a neutron star.  In addition, if we  consider the second nearest isolated compact star RXJ 0720.4-3125 (henceforth RXJ 0720), which has been detected by
ROSAT~\cite{haberl:1997,motch:1998} and observed with XMM-Newton \cite{cropper:2001,paerels:2001}, then we see  that the spectrum of RXJ 0720 is almost identical to
that of RXJ 1856. Indeed, RXJ 0720 exhibits a blackbody X-ray spectrum, a large X-ray to optical flux ratio, and  an optical emission in excess over the
extrapolated X-ray blackbody. In this case, however, the recent analysis of the optical, ultraviolet, and
X-ray data~\cite{kaplan:2003a} showed that the optical spectrum of RXJ 0720 is not well fitted by a Rayleigh-Jeans tail, but it is best fitted  by a non thermal power law. This strongly suggest that  the pulsar surface emission cannot account for the observed soft emission spectra, and this last emission must be  magnetospheric in origin. For instance, in Cea (2004b) we suggested that the soft spectrum originates from  synchrotron radiation~\cite{wallace:1977} emitted by electrons with power-law energy spectrum. In this case, the radiation in the soft spectrum should display a rather large linear polarization. \\
\indent
Our previous discussion lead to conclude that most realistic interpretation of the X-ray spectrum is thermal emission  due to the whole surface of the star.  Now, assuming  $d \; \simeq 180 \;  pc$, from Eq.~(\ref{2.24}) we get $ R^\infty \; \simeq \; 6.6 \, Km$. From this value of $ R^\infty$ we can solve Eq.~(\ref{2.25}) to constrain the mass and radius of RXJ 1856~\cite{cea:2004b}. The result of this analysis, displayed in   Fig.~\ref{fig1},  indicates  that  there are stable p-star configurations which agree with observational data.  However, it should be keep in mind that  $ R^\infty$ is a lower limit of  the true stellar radius. Nevertheless,  from Fig.~\ref{fig1} we see that the mass-radius curve for RXJ 1856 is more consistent with a soft equation of state. \\
\indent
Finally, from Fig.~\ref{fig1}  assuming that the observed compact objects are p-stars   we infer that the chromagnetic condensate
 is constrained in the rather narrow interval:
\begin{equation}
\label{2.29}
 \; 0.35  \; Gev \;  \lesssim \; \sqrt{gH} \;  \lesssim  \;  0.6  \; Gev  \; \; .
\end{equation}
As a consequence, the allowed  region for p-stars is the region in  Fig.~\ref{fig1} bounded by the two dot-dahed lines and the M-R curves with  $\sqrt{gH} = 0.6 Gev, 0.35 Gev$ .
\section{COOLING}
\label{cooling}
The thermal evolution of compact stars  is an important tool to investigate the state of dense matter at supra-nuclear densities. In fact, observations of the thermal photon flux emitted from the surface of the stars provide valuable information about the physical processes operating in the interior of these objects. \\
\indent
The cooling in p-stars has been discussed for the first time in Cea (2004a). As for neutron stars, the predominant cooling mechanism of newly formed p-stars is neutrino emission. In p-stars neutrino cooling dominates for about $10^2 - 10^3 \; years$. Subsequently, after the internal temperature has sufficiently dropped, the photon emission overtakes neutrinos. \\
\indent
Let us briefly review the cooling in p-stars~\cite{cea:2004a}.  Assuming  stars of uniform density and isothermal,  the cooling equation is:
\begin{equation}
\label{3.1}
 C_V \; \frac{d T}{d t}
\; = \; - \; (L_{\nu} \; + \; L_{\gamma}) \; ,
\end{equation}
where $L_{\nu}$ is the neutrino luminosity, $L_{\gamma}$ is the photon luminosity and $C_V$ is the specific heat. Assuming
blackbody photon emission from the surface at an effective surface temperature $T_S$ we get:
\begin{equation}
\label{3.2}
 L_{\gamma} \; = \; 4 \, \pi \, R^2 \, \sigma_{SB} \, T_S^4 \; ,
\end{equation}
where $\sigma_{SB}$ is the $Stefan-Boltzmann$ constant.  Following Shapiro \& Teukolsky (1983), in Cea (2004a) we assumed that the surface and interior temperature were related by:
\begin{equation}
\label{3.3}
 \frac{T_S}{T} \; = \; 10^{-2} \; a \; \; , \; \; 0.1 \; \lesssim a \; \lesssim 1.0
 \; .
\end{equation}
Eq.~(\ref{3.3}) is relevant for a p-star which is not bare, namely for p-stars which are endowed with a thin crust. It results~\cite{cea:2004b} that p-stars have a sharp edge of thickness of the order of about $ 1 \; fermi$. On the other hand, electrons which are bound by the coulomb attraction, extend several hundred $fermis$ beyond the edge. It follows, then, that on the surface of the star there is a positively charged layer which is able to support a thin crust of ordinary matter. The vacuum gap between the core and the crust of order of several hundred $fermis$  leads to a strong suppression of the surface temperature with respect to the core temperature. In principle, the actual relation between $T_S$ and $T$ can be obtained by  studying  the crust and core thermal interactions. In any case, our phenomenological relation Eq.~(\ref{3.3}) allows a wide variation of $T_S$, which encompasses the neutron star relation (see, for instance, Gudmundsson 1983). Moreover, as we discuss below, our cooling curves display a rather weak dependence on the parameter $a$ in Eq.~(\ref{3.3}). \\
\indent
In Cea (2004a) we showed that the dominant cooling processes by neutrino emission are the direct $\beta$-decay quark reactions~\cite{iwamoto:1980,burrows:1980}:
\begin{equation}
\label{3.4}
 d \; \rightarrow \; u \, + \, e \, + \, \overline{\nu}_e \; \; \; , \; \; \;
u \, + \, e \, \; \rightarrow \; d \,  + \, \nu_e \;  \; \; .
\end{equation}
We find the following neutrino luminosity~\cite{cea:2004a}:
\begin{equation}
\label{3.5}
 L_{\nu} \; \simeq \; 3.18 \; 10^{36} \; \frac{erg}{s} \; T_9^8 \;
 \frac{M}{M_{\bigodot}} \; \frac{\varepsilon_0}{\varepsilon} \;
   \frac{\sqrt{gH}}{1 \, GeV} \;  \;  ,
\end{equation}
where $T_9$ is the temperature in units of $10^9$ $\, {}^\circ K$, and $\varepsilon_0 \, \simeq  2.51 \; 10^{14} g \, cm^{-3}$ is the nuclear density.  So that, for typical  values of parameters we have:
\begin{equation}
\label{3.6}
 L_{\nu} \; \sim \;  10^{36} \; \frac{erg}{s} \; T_9^8 \;  \;  \;  .
\end{equation}
Note that the neutrino luminosity $L_{\nu}$ has the same temperature dependence as the neutrino luminosity by  modified
URCA reactions in neutron stars (see, for instance Shapiro \& Teukolsky  1983), but it is more than two order of magnitude smaller. This peculiar dependence on the core temperature is due to the presence of the strong chromomagnetic condensate which strongly constraints the quark transverse motion. \\
\indent
It is worthwhile to stress that, since our neutrino luminosity is reduced by more than two order of magnitude with respect to neutron stars,
the maximum allowed quiescent luminosity of isolated pulsars is about two order of magnitude greater than the maximum allowed surface luminosity in neutron stars \cite{VanRiper:1991}. Thus, while our theory allows to account for luminosities  up to $10^{36} \;
erg \, s^{-1}$ as observed in magnetars~\cite{cea:2006}, the standard model based on neutron stars is in striking  contradiction with observations. \\
\indent
A further support on our neutrino luminosity Eq.~(\ref{3.6}) comes from superbursts, namely rare, extremely energetic, and long duration type I X-ray bursts, from low mass X-ray binaries (see  Kuulkers  2004). Indeed, it has been pointed out that to account for the observed ignition depths in superbursts one needs slow neutrino emission processes  with emissivity  \cite{Page:2005,Stejner:2006}: 
\begin{equation}
\label{3.7}
 \epsilon_{\nu} \; =  \;  Q_{\nu}  \;  \; T_9^8 \;  \frac{erg}{cm^{3} \, s} \;  \;  , \;  \;  Q_{\nu} \; \sim \; 10^{18} \; - \; 10^{22} .
\end{equation}
It is evident that the phenomenological neutrino emissivity Eq.~(\ref{3.7}) needed to explain observed superbursts from low mass X-ray binaries cannot be accounted for within the standard neutron star model. On the contrary, from our Eq.~(\ref{3.6}) we infer for the neutrino emissivity in p-stars:
\begin{equation} 
\label{3.8}
 \epsilon_{\nu} \; \sim  \; 10^{18}   \;  \; T_9^8 \;  \frac{erg}{cm^{3} \, s} \;  \;  , 
\end{equation}
which naturally fits in the  phenomenological allowed range of values. \\
\begin{figure}[ht]
   \centering
   \epsscale{0.56}
  \plotone{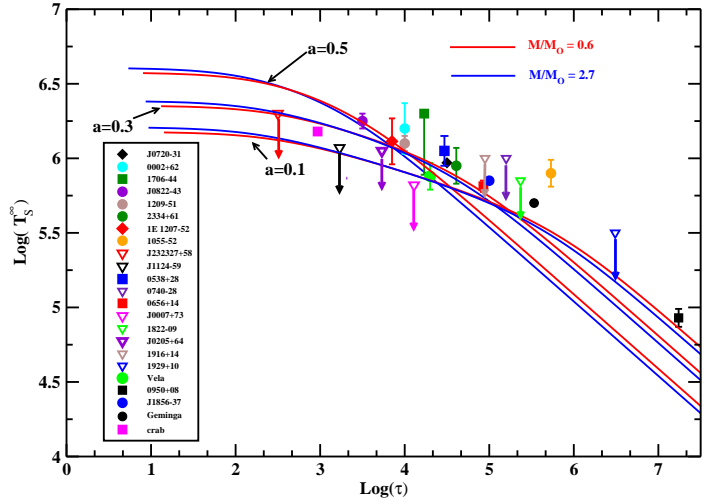}
\caption{\label{fig2}
Comparison of p-star cooling curves with several data for the effective surface
temperature. Different lines correspond to  compact star mass  values indicated in the legend
and  different values of the parameter $a$ in Eq.~(\ref{3.3}).  Data points with error bars  are taken from Yakovlev \& Pethick (2004) and references therein.}
\end{figure}
\indent

Finally, to determine the  thermal evolution of p-stars we need the specific heat \cite{cea:2004a}:
\begin{equation}
\label{3.9}
 C_V \; \simeq \; 0.92 \; 10^{55} \;  T_9 \;
 \frac{M}{M_{\bigodot}} \; \frac{\varepsilon_0}{\varepsilon} \;
   (\frac{\sqrt{gH}}{1 \, GeV})^2 \; ,
\end{equation}
which, for typical parameter values, in physical units reads:
\begin{equation}
\label{3.10}
 C_V \; \sim   \; 10^{39} \; \frac{erg}{{}^\circ K} \: T_9  \;  \; .
\end{equation}
Note that, from Eq.~(\ref{3.10}) it follows that the p-star specific heat is of the same order of the neutron star specific heat \cite{shapiro:1983}. To obtain the thermal history of a p-star we integrate Eq.~(\ref{3.1}) by assuming the initial temperature  $T^{(i)}_9 = 1.4$ \cite{cea:2004a}. We stress, however, that the thermal history is almost independent on the assumed initial temperature as long as $T^{(i)} \; \leq  \; 10^{10}\; {}^\circ K$. \\
\indent
In Fig.~\ref{fig2} we report our cooling curves for three different values of the parameter $a$ in Eq.~(\ref{3.3}). It is worthwhile to note that the cooling curves are almost independent on the p-star mass. Moreover, there is a weak dependence on $a$ up to age $ \tau \, \sim \, 10^3 \, years $. We
compare our theoretical cooling curves with several pulsar data taken from the literature.  \\
\indent
In any case we see that the agreement between theoretical cooling curves and observational data is quite satisfying. In particular, we see that, at variance with neutron stars,  our peculiar neutrino luminosity allows sizeable surface effective temperature, up to $10^5 \, {}^\circ K$, for compact  stars with age $\tau > 10^6 \, years$. Indeed,  Fig.~\ref{fig2}  suggests that this distinguishable feature of the p-star model is corroborated by observations.
\section{CONCLUSIONS}
\label{conclusion}
The proposal for p-stars originates from our non-perturbative investigations of QCD (Cea \& Cosmai 2003, 2005) which suggested that quarks and gluons get deconfined in strong enough chromomagnetic fields. This, in turns, leads us to argue that the deconfined QCD vacuum is characterised by long-range chromomagnetic correlations  and that p-matter, namely almost massless up and down quarks immersed in a chromomagnetic condensate, is formed in the collapse of the core of an evolved massive star~\cite{cea:2004a}. In addition. we already argued that p-stars are more stable than neutron stars whatever the value of the chromomagnetic condensate.  As a consequence, the true ground state of QCD in strong gravitational fields is not hadronic matter, but p-matter. In other words, the final collapse of massive stars leads inevitably to the formation of p-stars. \\
\indent
The results  discussed in the present paper, indeed, indicate that p-stars are  able to reproduce in a natural manner several
observations from pulsars. On the other hand, we have seen that  within the standard model based on neutron stars there is some tension in the  determination of the  equation of state for  high density neutron matter. However, we would stress that mass-radius constraints are affected by important systematic errors and are not therefore so constraining. Thus, we need more precise mass-radius determinations to reach a firm conclusion.
\end{document}